# A Topology-Aware Neural Network for Fault Location With Missing Measurements

Fan Jiang
Department of Electrical and Computer Engineering
University of Houston
Houston, TX, USA
fjiang6@uh.edu

Linhan Fang
Department of Electrical and Computer Engineering
University of Houston
Houston, TX, USA
lfang7@cougarnet.uh.edu

Xingpeng Li
Department of Electrical and Computer Engineering
University of Houston
Houston, TX, USA
xli83@central.uh.edu

Kaushik Rajashekara
Department of Electrical and Computer Engineering
University of Houston
Houston, TX, USA
ksraja@uh.edu

***Abstract*—This paper proposes a topology-aware neural network-based fault locator (TANFL) for radial distribution feeders, which uses voltage and current phasors obtained after the decay of fault induced transients, as well as feeder connection relationships and line parameters, as inputs. The shared bus encoder, bus readout module, and gated line readout module work together to identify the faulty bus or line and estimate the continuous fault location on the selected line. During training, explicit missing data indicators and random bus data block masking are introduced, enabling a single trained model to handle both complete measurement sets and scenarios where any permitted bus's measurement device fails. On the IEEE 15-bus feeder, this paper compares TANFL with a conventional multilayer perceptron (MLP) and the formula-based localization method under identical data, measurement availability scenarios, and evaluation metrics. On the test partition, TANFL outperforms the conventional MLP; with complete measurements, its line fault location performance is slightly lower than that of the formula-based method, but in the average results for scenarios with one missing bus measurement block, its accuracy is the highest among the three methods at a 5% tolerance.**

***Keywords—distribution systems, fault location, missing measurements, neural networks, topology-aware learning.***

## I. INTRODUCTION

Single-line-to-ground (SLG) faults frequently occur on distribution feeders due to vegetation, weather, equipment failures, and animal contact. After a protective device trips, the utility must first locate the damaged section before maintenance personnel can isolate it and restore power. Fault location on distribution feeders is more challenging than on transmission lines: feeders have numerous branches and tap connected loads, inconsistent line parameters, continuously varying loads, and fault resistances that are typically high and unknown [1], [2]. Furthermore, a single set of measurements may correspond to multiple locations with similar electrical characteristics on different branches; therefore, the location result must also identify the component affected by the fault. The increasing deployment of phasor measurement units and smart electronic devices at the distribution level has created the conditions for multipoint synchronous phasor measurement [3], [4], but in practice, it cannot be guaranteed that all measuring devices will always be available.

Physics-based methods utilize circuit relationships and measured voltages and currents to infer the fault location. Liao [2] derived a location equation based on local substation measurements and a known network impedance model. Lotfifard et al. [1] combined substation records with voltage dip measurements at selected downstream locations and pre-fault load model updates. Majidi and Etezadi-Amoli [3] performed location estimation using a small number of busbar voltages measured before and during the fault, in conjunction with line impedance models and least-squares estimation. Jia et al. [4] identified the faulted section based on sparse voltage measurements and a network admittance model. Other physics-based approaches include the Takagi method, which estimates the fault distance using single ended fundamental phasors [5], and its extended methods further account for branch circuits, tap loads, and unbalanced distribution feeders [6], [7]. These methods have clear physical interpretations and achieve high accuracy when their assumptions hold. However, they may yield multiple electrically equivalent fault locations; and if any required measurement is missing, the corresponding formulas cannot be calculated.

Learning-based methods utilize simulated or recorded fault data to learn the mapping from measured quantities to fault locations. Thukaram et al. [8] combined support vector classification with a feedforward neural network to estimate fault locations; Liang et al. [9] used an adaptive convolutional network to select the faulty line, and then employed a two-terminal phasor to calculate the fault distance. Research has also been conducted on estimating continuous positions on a line using topology-aware networks. Sun et al. [10] encoded fault distance into the outputs of the nodes at both ends of the line in a graph neural network and used this to recover the relative position within the line; Khodayar et al. [11] and Saffari et al. [12] utilized dynamic graph models and spatio-temporal graph models to estimate normalized fault distance and classify fault types. The issue of missing measurements has also received attention. Wang and Hu [13] performed regression on the relative fault location and tested scenarios where node measurements were randomly removed; however, they did not consider busbar faults. Savaei et al. [14] combined decision tree line selection with a multilayer perceptron (MLP) distance regression based on traveling wave arrival times in a transmission network and investigated scenarios where one to three busbar records were missing. Wu et al. [15] fused substation alarms, SCADA, and waveform data to classify locations, including busbars and lines, under conditions of missing channel data; however, they did not output continuous locations.

Overall, the aforementioned studies cover the three capabilities in various combinations: continuous location regression, differentiation between busbar and line elements, and robustness to missing inputs, none has simultaneously possessed all three capabilities within a single model. This paper proposes the topology-aware neural network-based fault locator (TANFL), which combines these three capabilities by taking as inputs voltage and current phasors obtained after the fault persists but the transient process has decayed (hereafter referred to as post-transient phasors), known feeder connection relationships, and line parameters. The main contributions are as follows:

- Proposed a single locator that identifies the faulted component in a joint space of bus classes and line classes and, for a line fault, estimates the continuous position along the line.
- Combined a missing indicator and random bus block masking under complete and missing measurements to support component identification and position estimation with one missing bus measurement block.
- Compared with a conventional MLP and a formula-based locator adapted from [16] under identical data, measurement scenarios, and metrics to identify each method's limitations.

## II. Topology-Aware Neural Network-Based Fault Locator

### A. *Structure of the TANFL*

Fig. 1 illustrates the proposed TANFL and its components.

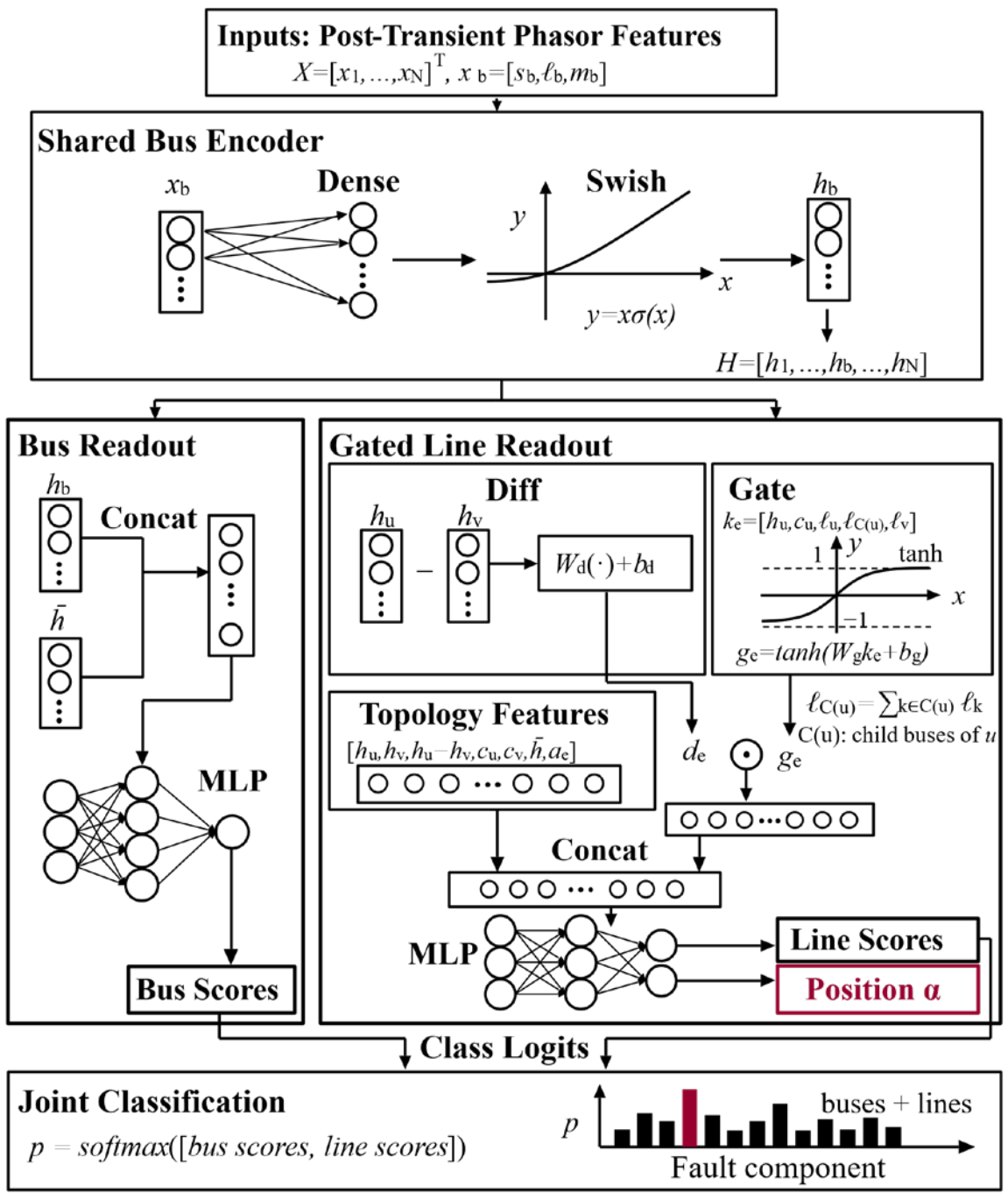

Fig. 1. TANFL architecture.

*1) Inputs*: At bus *b*, the input vector collects the preprocessed features of that bus, computed from the post-transient phasors:

$$x_b = [s_b, \ell_b, m_b] \in \mathbb{R}^n \tag{1}$$

where $s_b \in \mathbb{R}^{2q}$ contains their scaled real and imaginary parts, $q$ denotes the number of phasors measured at a bus, $\ell_b \in \mathbb{R}^q$ contains their scaled log magnitudes, and $m_b$ is the missing indicator, which equals 0 for an available block and 1 for a missing block. When a bus measurement block is missing, all corresponding measurement features are set to zero, and the missing indicator is set to 1. The number of features per bus is $n = 3q + 1$. The network input is the matrix $X$ obtained by stacking the vectors of all buses:

$$X = [x_1, \dots, x_N]^{\mathrm{T}} \in \mathbb{R}^{\mathrm{N} \times n} \tag{2}$$

where $N$ is the number of buses.

*2) Shared Bus Encoder:* A dense layer with swish activation, shared by all buses, maps each $x_b$ to a bus embedding $h_b \in \mathbb{R}^d$, where $d$ is the embedding dimension. Sharing the weights keeps the encoder independent of the number of buses. $H = [h_1, \dots, h_N]$ collects the embeddings of all buses.

*3) Bus Readout:* This component scores the likelihood that the fault is at each bus. The mean embedding $\bar{h} = N^{-1}\Sigma_b\, h_b$ summarizes the whole feeder. For every bus $b$, it concatenates $[h_b, \bar{h}]$, applies a dense layer of $d$ swish units, and outputs one bus score.

*4) Gated Line Readout:* This component scores the likelihood that the fault is on each line and estimates the position of the fault along that line. For line $e = (u, v)$, $u$ is the upstream bus and $v$ is the downstream bus. The topology features path concatenates $h_u$, $h_v$, their difference $h_u - h_v$, the sums $c_u$ and $c_v$ of the embeddings of the child buses of $u$ and $v$, the mean embedding $\bar{h}$, and the standardized resistance and reactance $a_e$ of the line. This path exposes the local connections and the downstream information of every line without a separate encoder for each bus.

The projected difference and the learned gate are computed as follows:

$$d_e = W_d(h_u - h_v) + b_d \tag{3}$$

$$g_e = \tanh(W_g\, k_e + b_g) \tag{4}$$

where $W_d$ and $b_d$ are the weight matrix and the bias of the diff path, $W_g$ and $b_g$ are the weight matrix and the bias of the gate path, and the gate input $k_e$ collects the node and log features of the line: $h_u$, $c_u$, the scaled log magnitudes $\ell_u$ and $\ell_v$, and the sum of $\ell_k$ over the child buses $k$ of $u$.

The elementwise product $d_e \odot g_e$, where $\odot$ denotes elementwise multiplication, is concatenated with the topology features; an MLP with one hidden layer of $d$ swish units then maps this vector to one line score and one position $\alpha_e$.

*5) Joint Classification:* The bus scores and line scores form the class logits, and the probability vector is given by

$$p = \mathrm{softmax}([\text{bus scores}, \text{line scores}]) \tag{5}$$

Its entry $p_c$ is the probability that the fault is at component $c$, and the selected component is

$$\hat{c} = \arg\max_c p_c \tag{6}$$

If $\hat{c}$ is a line, the final position estimate is obtained by selecting that line's output and clipping it to the interval [0, 1]:

$$\hat{\alpha} = \min\{1, \max\{0, \alpha_{\hat{c}}\}\} \tag{7}$$

where $\alpha_{\hat{c}}$ is the component of the position vector corresponding to the selected line $\hat{c}$, and $\alpha = 0$ and $\alpha = 1$ denote the upstream and downstream ends of the line.

Direct bus faults have their own classes and need no position. The network output collects the probability vector $p$ and the vector $\alpha$ of the unclipped positions of all lines, with one component per line:

$$y = [p, \alpha], p \in \mathbb{R}^{N+K}, \alpha \in \mathbb{R}^K \tag{8}$$

where $y$ is the estimated location and $K$ is the number of lines.

## B. Data Generation and Preprocessing

The dataset was obtained through simulations in PSCAD/EMTDC, with each bus equipped with three-phase voltage and incoming current measurement channels. For each case, a faulted line and its location were selected, the fault resistance was sampled from a log-uniform distribution, and a multiplier was independently sampled for each load. This paper considers SLG faults, with fault locations covering the whole line and busbar faults. After the DC component caused by the fault has sufficiently decayed, the post-transient phasors are extracted from the recorded AC fundamental components of the voltage and current signals.

All phasors are rotated using the phase A voltage of the power supply bus as the phase reference, and then scaled using constants obtained from fitting the training partition. The real and imaginary parts of the voltage are first centered and then divided by the same voltage scaling factor. The current components are first centered, then compressed using the asinh function to limit the dynamic range of high current cases, and subsequently divided by the same current scaling factor.

## C. Training and Validation Procedure

After obtaining the dataset, it is first divided into training, validation, and test partitions, and any preprocessing constants are then fitted to ensure that the scaling statistics are derived solely from the training partition.

The loss function in (9) consists of a classification term, a network path term, and a location term. The path distance is calculated by summing the series impedance magnitudes of the line portions along the feeder path connecting the predicted and true fault locations. If the path traverses only part of a line, only that part contributes to the distance. The path loss in (10) weights the normalized distances of the candidate locations by their predicted probabilities. The location loss in (11) measures the absolute error in the estimated position on the true faulted line.

$$L = 0.25\, L_{CE} + L_{path} + L_{\alpha} \tag{9}$$

$$L_{\text{path}} = \sum_c p_c \frac{d_G(y_c, y^*)}{\text{median}_e |Z_e|} \tag{10}$$

$$L_{\alpha} = |\alpha_{e^*} - \alpha^*| \tag{11}$$

where $L_{\text{CE}}$ is the categorical cross entropy over the bus and line classes. $d_{\text{G}}$ is the shortest path distance along the feeder when each line $e$ has weight $|Z_{\text{e}}|$. $Z_{\text{e}}$ is the series impedance of line $e$. $y^*$ is the true fault location, $y_c$ is the location associated with candidate class $c$, and $p_c$ is its probability. $e^*$ is the true faulted line, $\alpha^*$ is the true position on that line, and $\alpha_{e^*}$ is the position output of that line without clipping.

This combination penalizes large errors in electrical distance while providing a direct regression target, thereby avoiding sigmoid saturation issues at both ends of the line. During training, a measurement from a bus where missing values are allowed is removed from each sample with a probability of 0.5, ensuring that the network is exposed to an equal proportion of complete and incomplete inputs. The path mean absolute errors (MAEs) are calculated on the validation partition for both complete input and scenarios with one missing bus measurement block.

After TANFL training is complete, evaluation is performed on the test partition of the dataset. The test partition is evaluated under conditions of complete measurements and by sequentially removing data blocks from each bus where missing measurements are permitted.

# III. Case Study

This section compares the trained TANFL with a conventional MLP and a formula-based locator under complete measurements and scenarios with one missing bus measurement block.

## A. Test System

The example uses an IEEE 15-bus feeder. This feeder consists of 15 buses and 14 lines, as shown in Fig. 2, with a rated voltage of 13.8 kV. Bus 1 is connected to the voltage source. For the remaining buses, current measurements are taken at the receiving end of the upstream lines. Each bus therefore provides $q$ = 6 phasors, including the three-phase voltages and currents, and one missing indicator, resulting in $n$ = 19 features per bus. Buses 1–4 remain available in all scenarios. In missing data tests, one complete block of voltage and current data from buses 5–15 is removed.

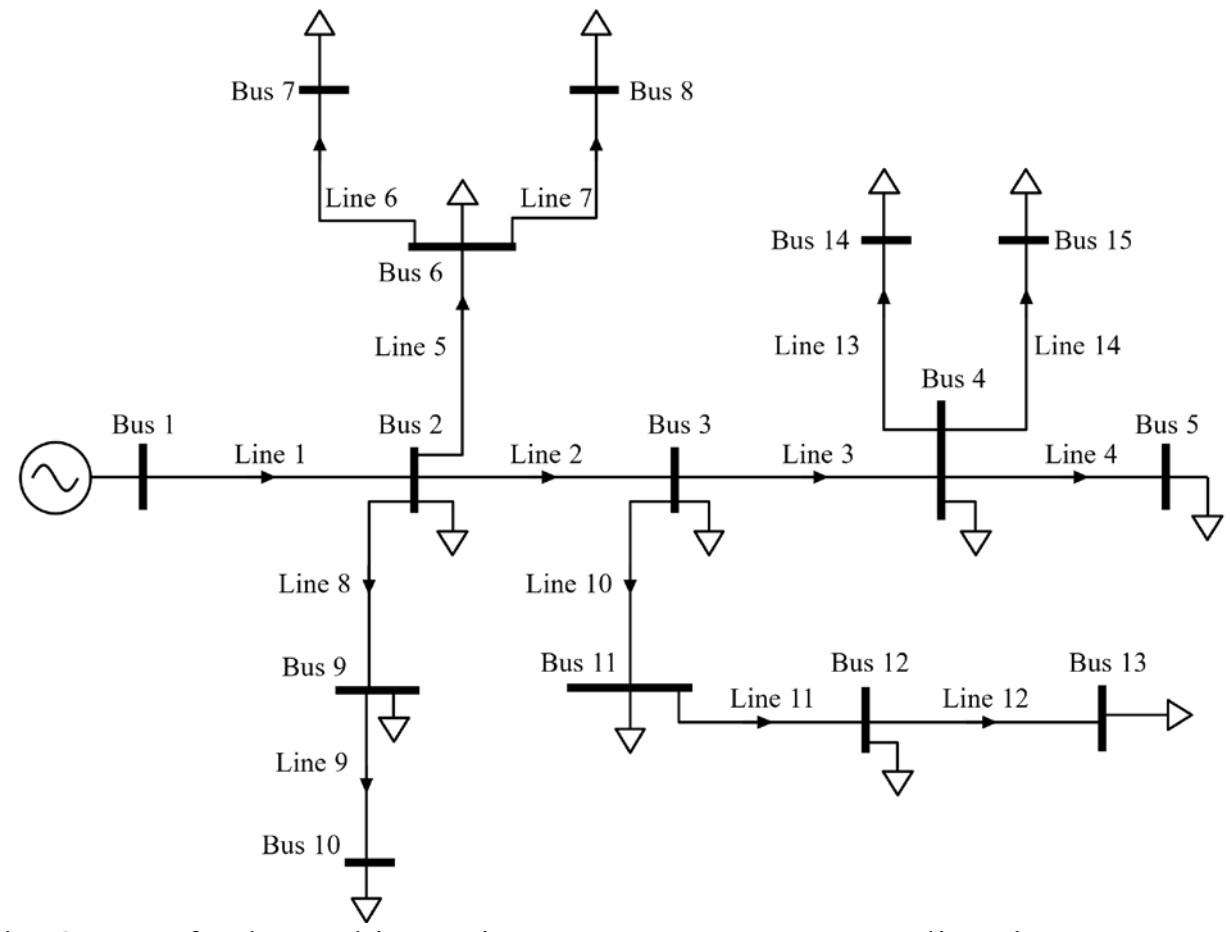

Fig. 2. Test feeder and incoming current measurement directions.

*1) Dataset Acquisition:* Only phase A to ground faults were simulated on lines and buses. Loads were modeled as constant impedance loads from the feeder model, with the rated power scaled according to the corresponding load factor for each case. Fault resistance was sampled on a logarithmic scale within the range of 0.01–20 Ω, and each load factor varied independently within the range of 80%–120%. For each line, 110 line fault cases were set up, distributed along the entire length of the line; for each bus, 20 direct bus fault cases were set up. The resulting 1,840 development cases were divided into training, validation, and test partitions in the ratio of approximately 70%, 15%, and 15%, respectively, with each bus appearing in all partitions.

*2) Training settings:* The embedding dimension $d$ is set to 32, and TANFL contains 13,795 trainable parameters. Training uses the Adam optimizer with an initial learning rate of $10^{-3}$, a batch size of 64, and a maximum of 300 training epochs. The checkpoint with lowest validation loss is saved.

*3) Comparison of localization methods:* The two baseline methods receive the same post-transient phasors and are evaluated using the same metrics as the proposed TANFL. The conventional MLP flattens the input matrix $X$ into a vector and passes it through two fully connected hidden layers containing 128 and 64 swish units, respectively. The outputs are identical to that of TANFL. This model contains 47,659 parameters, which is 3.5 times that of TANFL, and is trained using the loss function in (9) along with the same rules.

The formula-based localization method also utilizes post-transient voltage and current phasors, line impedance, and feeder connection relationships. Position estimation requires measurements at both ends of the line. The position calculation is adapted from [16], while the busbar and line selection rules are tailored to this specific feeder. Among the busbars with the required measurements, the busbar with the largest current balance residual is selected, and its feeders are evaluated. Next, the line with the largest estimated position is selected. If the estimated position exceeds the threshold of 2% of the line length, it is classified as a line fault, and the position is truncated to [0, 1]; otherwise, or if there are no outgoing lines, it is classified as a busbar fault. Candidate elements lacking the necessary measurements are excluded.

*4) Evaluation metrics:* A line fault estimate is counted as correct at tolerance $\tau$ only when the predicted line matches the true faulted line and the position error on that line is within the tolerance, that is, when

$$\begin{cases} \hat{c} = e^* \\ |\hat{\alpha} - \alpha^*| \leq \tau \end{cases} \tag{12}$$

where $\alpha^*$ is the true position and $\tau$ is the tolerance.

A busbar fault is counted as a hit when the predicted busbar is correct, or when the predicted location is on a line connected to that busbar and is no more than $\tau$ away from it. The accuracy rate is reported for $\tau$ = 1%, 3%, 5%, and 10%. Path MAE is the average of path distance $d_G(\hat{y}, y^*)$ across all evaluation cases. The definition of path distance is provided in Section II.C. A smaller path MAE indicates a shorter average electrical distance between the predicted location and the true location. This metric is measured in terms of the amplitude normalized series impedance relative to the system baseline.

## B. Localization with Complete Measurements

Under full measurement conditions, Fig. 3(a) and 3(c) show the line and busbar accuracy rates for the same test set at tolerances of 1%, 3%, 5%, and 10%. Table I reports the line and busbar accuracy rates and path MAE for the test partition of the development dataset at a 5% tolerance.

As shown in Fig. 3(a) and (c), TANFL achieves accuracy comparable to the formula-based locator at the 5% and 10% tolerances under complete measurements, and both outperform the conventional MLP for line and bus faults. TANFL's advantage over the MLP is consistent with its explicit use of feeder connectivity, line parameters, and features from neighboring buses, which provide structural information for identifying the faulted component and estimating its position. The formula-based locator benefits from noise-free simulated phasors and line parameters that match the simulation, achieving a mean position error of only 0.02% of the line length on correctly identified lines. Its remaining line fault misses occur near the line start, where the 2% decision threshold causes classification as upstream bus faults. These cases remain incorrect regardless of the evaluation tolerance. In contrast, most TANFL misses at tight tolerances arise from position errors on the correct line, so its accuracy increases as the tolerance is relaxed.

Table I quantifies the comparison at a 5% tolerance. TANFL correctly locates 97.48% of line faults and 100% of bus faults, compared with 80.67% and 93.33% for the conventional MLP. Its line fault accuracy is only 0.42 percentage points below the formula-based locator's 97.90%, while both achieve 100% bus fault accuracy. TANFL also reduces the path MAE from the MLP's 3.21 to 1.37 × $10^{-3}$ p.u., indicating a smaller average electrical distance error. The formula-based locator achieves the lowest path MAE, 0.05 × $10^{-3}$ p.u., under these simulation conditions.

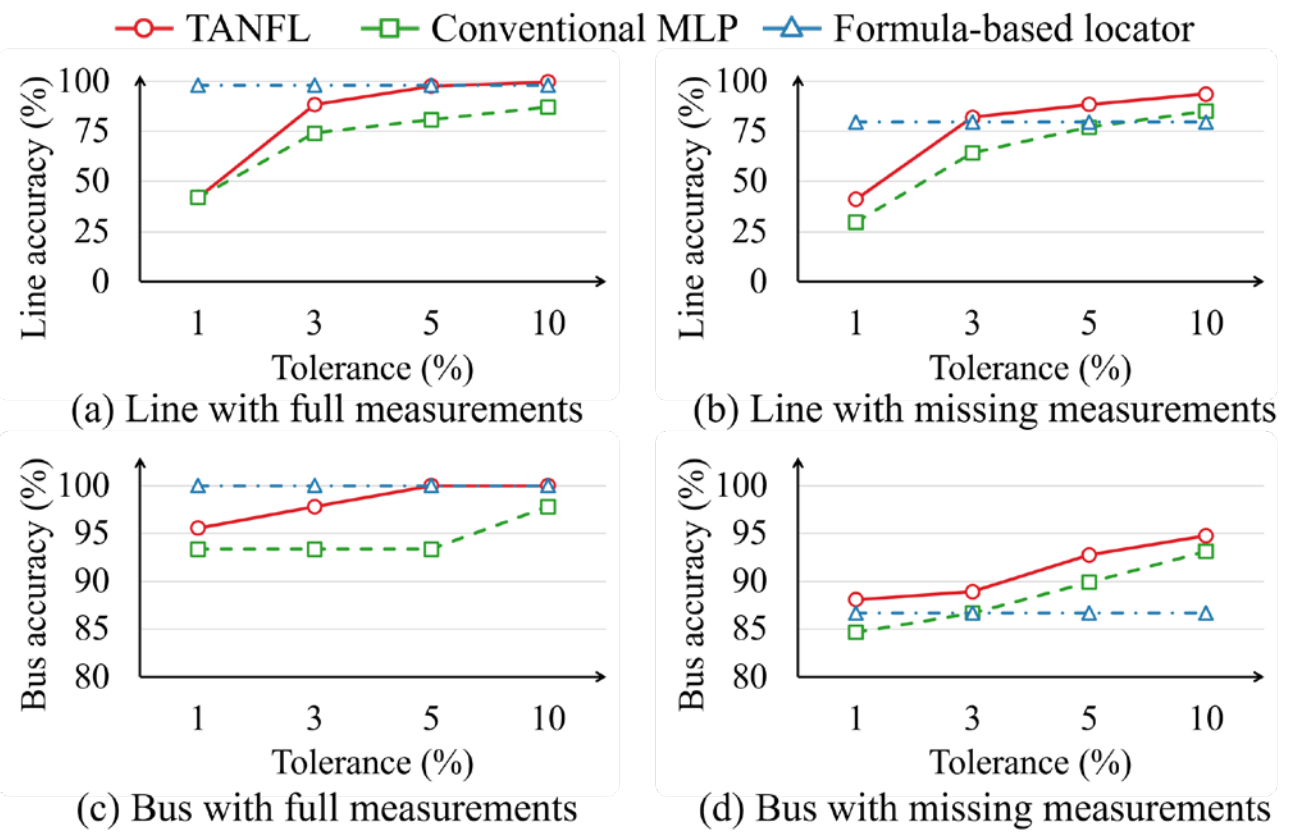

Fig. 3. Accuracy on the test partition at 1%, 3%, 5%, and 10% tolerance: (a) line faults and (c) bus faults with complete measurements; (b) line faults and (d) bus faults with one missing bus measurement block.

TABLE I. 5% TOLERANCE ACCURACY AND PATH MAE WITH COMPLETE MEASUREMENTS

| Method | Line 5% (%) | Bus 5% (%) | Path MAE ($10^{-3}$ p.u.) |
|---|---|---|---|
| TANFL | 97.48 | 100.00 | 1.37 |
| Conventional MLP | 80.67 | 93.33 | 3.21 |
| Formula-based locator | 97.90 | 100.00 | 0.05 |

## C. Localization with Missing Measurements

Each time a complete bus data block was removed from buses 5–15, the missing bus was marked via an indicator channel, and the network was evaluated without retraining. Fig. 3(b) and 3(d) show the average line and bus accuracy rates at four tolerance levels. Table II reports the average line and bus accuracy rates, as well as path MAE, for the test partitions at a 5% tolerance.

TABLE II. 5% TOLERANCE ACCURACY AND PATH MAE WITH ONE MISSING BUS MEASUREMENT BLOCK

| Method | Line 5% (%) | Bus 5% (%) | Path MAE ($10^{-3}$ p.u.) | True element computable (%)* |
|---|---|---|---|---|
| TANFL | 88.35 | 92.73 | 2.80 | 100 |
| Conventional MLP | 76.89 | 89.90 | 4.33 | 100 |
| Formula-based locator | 79.49 | 86.67 | 49.03 | 82.04 |

* True element computable is the percentage of cases where available measurements allow evaluation of the true faulted element.

Within each scenario, line and bus accuracy were calculated using all cases corresponding to the respective fault types, while path MAE and the percentage of cases in which the true faulted element could be evaluated were computed using all evaluation cases. All averages reported under missing data conditions were first calculated for each scenario and then equally weighted averaged across the eleven scenarios.

Fig. 3(b) and (d) show the accuracy under missing measurements, averaged equally over the eleven scenarios with one missing bus measurement block on the test partition. TANFL outperforms the conventional MLP at all four tolerances for both line and bus faults. For line faults, TANFL achieves accuracies of 41.14%, 81.82%, 88.35%, and 93.39% at the 1%, 3%, 5%, and 10% tolerances, respectively. The formula-based locator leads only at 1%, with an accuracy of 79.49% that remains almost unchanged across the four tolerances. For bus faults, TANFL also exceeds the formula-based locator at every tolerance, including 88.08% versus 86.67% at 1%. TANFL's performance is consistent with its use of a missing indicator and bus block masking during training, which allow it to use the remaining measurements when a block is unavailable. The formula-based locator, however, requires specific currents to calculate the current balance residual and both terminal voltages to estimate the line position. Missing measurements therefore exclude some candidate elements. Its mean position error on correctly identified lines remains only 0.02% of the line length, whereas incorrect line selections cannot be recovered by relaxing the evaluation tolerance.

Table II summarizes the results at a 5% tolerance. TANFL achieves line and bus fault accuracies of 88.35% and 92.73%, compared with 76.89% and 89.90% for the conventional MLP and 79.49% and 86.67% for the formula-based locator. TANFL also has the lowest path MAE, at $2.80 \times 10^{-3}$ p.u., compared with $4.33 \times 10^{-3}$ p.u. for the MLP and $49.03 \times 10^{-3}$ p.u. for the formula-based locator. The true faulted element remains computable in only 82.04% of cases for the formula-based locator; this percentage describes more about computational availability than accuracy. With Bus 6 missing, TANFL's 5% line fault accuracy is 79.83%, its lowest value among the eleven scenarios.

Overall, under conditions of complete measurements in the test partition, the formula-based localization method achieved the highest accuracy among the three methods for line faults at tolerance levels of 1%, 3%, and 5%; TANFL lagged by less than 0.5 percentage points at the 5% tolerance level and led at the 10% tolerance level, with both methods successfully locating all busbar faults at the 5% tolerance level. After averaging the results for the eleven scenarios with one missing measurement block on the test partition, TANFL achieved the highest accuracy for line faults at 3%, 5%, and 10% tolerance levels, as well as for all tolerance levels of busbar faults; the formula-based localization method led only at the 1% tolerance level for line faults. These results demonstrate the advantage of the proposed TANFL under the tested missing measurement scenarios, while maintaining competitive accuracy with complete measurements.

## IV. Conclusion

This paper presented TANFL, a topology-aware neural fault locator that uses post-transient phasors to identify the faulted bus or line and estimate the position along the selected line, while accommodating one missing bus measurement block. On the IEEE 15-bus feeder, TANFL achieved higher accuracy than the conventional MLP at a 5% tolerance with fewer parameters under both complete and missing measurements. With complete measurements, its line fault accuracy was comparable to that of the formula-based locator at the 5% and 10% tolerances. Averaged over the eleven tested scenarios with one missing bus measurement block, TANFL achieved the highest line fault accuracy at the 3%, 5%, and 10% tolerances and the highest bus fault accuracy at all tested tolerances. These findings support its effectiveness under the tested measurement conditions. Future work will extend the method to other fault types.